\documentclass[twocolumn,amsmath,amssymb,longbibliography,aps, prb,floatfix]{revtex4-2}
\pdfoutput=1
\usepackage{hyperref}
\usepackage{url}
\usepackage{color}
\usepackage{acro}
\usepackage{chemmacros}
\usepackage[normalem]{ulem}

\newcommand{\beq}{\begin{eqnarray}}
\newcommand{\eeq}{\end{eqnarray}}

\usepackage{graphicx}

\newcommand\numberthis{\addtocounter{equation}{1}\tag{\theequation}}

\def\up{\uparrow}

\usepackage{bbm}
\usepackage{diagbox}

\begin{document}
\title{Spin fluctuations in the ultranodal superconducting state of Fe(Se,S)}
\author{Yifu Cao$^1$, Chandan Setty$^2$, Andreas Kreisel$^3$, Laura Fanfarillo$^{1,4}$ and P.J. Hirschfeld$^1$}
\affiliation{$^1$Department of Physics, University of Florida, Gainesville, Florida 32603, USA}

\affiliation{$^2$Department of Physics and Astronomy, Rice Center for Quantum Materials, Rice University, Houston, Texas 77005, USA}

\affiliation{$^3$Niels Bohr Institute, University of Copenhagen, Jagtvej 155 , DK-2200, Copenhagen, Denmark}

\affiliation{$^4$Istituto dei Sistemi Complessi (ISC-CNR), Via dei Taurini 19, I-00185 Rome, Italy}

\begin{abstract}
The iron-based superconductor FeSe isovalently substituted with S displays an abundance of remarkable phenomena that have not been fully understood, at the center of which are apparent zero-energy excitations in the superconducting state in the tetragonal phase. The phenomenology has been generally consistent with the proposal of the so-called ultranodal states where Bogoliubov Fermi surfaces are present. Recently, nuclear magnetic resonance measurements have seen unusually large upturns in the relaxation rate as temperature decreases to nearly zero in these systems, calling for theoretical investigations. In this paper, we calculate the spin susceptibility of an ultranodal superconductor including correlation effects within the random phase approximation. Although the non-interacting mean-field calculation rarely gives an upturn in the low temperature relaxation rate within our model, we found that correlation strongly enhances scattering between coherent parts of the Bogoliubov Fermi surface, resulting in robust upturns when the interaction is strong. Our results suggest that in addition to the presence of Bogoliubov Fermi surfaces, correlation and multiband physics also play important roles in the system's low energy excitations.

\end{abstract}
\maketitle

\section{Introduction}

Iron-based superconductors have been drawing lots of research interest for almost two decades since their discovery, for their relatively high $T_c$,  simple structure and the interplay between rich phenomena including nematicity, magnetization and non-trivial topology\cite{Fernandes_Nature2022}. Among all families of iron-based superconductors, the chalcogenide 11 material FeSe has a distinct phase diagram where a nematic transition can occur without accompanying magnetic order\cite{Coldea2019,Kreisel2020,Sun2016,Matsuura2017,Bendele2010,Terashima2015}. The parent compound FeSe shows a nematic transition at around $90$K, a superconducting (SC) transition at $9$K and no magnetic order under ambient pressure. Upon applying hydrostatic pressure the nematicity is suppressed and antiferromagnetic (AFM) order develops. The AFM order in FeSe under pressure should resemble that observed in iron pnictides, and is likely a stripe order with in-plane magnetic moments \cite{Fernandes_Nature2022,Wang2016,delaCruz2008,Stadel2022}. 

On the other hand, the S-substituted FeSe does not show strong evidence for long-ranged magnetic order, but exhibits peculiar changes in its superconducting states across the nematic quantum critical point at around $0.17$ sulfur substitution. For $x>0.17$,  the normal state of FeSe$_{1-x}$S$_x$ is tetragonal, established by various measurements of the electronic structure\cite{Coldea2021,Reiss2017,Coldea2019}; and the transport properties show non-Fermi liquid behavior near the quantum critical point\cite{Licciardello2019}. The superconducting state shows curiously large zero energy density of states (DOS), which has so far been evidenced by specific heat and thermal transport measurements\cite{Sato2018}, scanning tunneling microscopy (STM)\cite{Hanaguri2018}, angular-resolved photoemission spectroscopy (ARPES)\cite{nagashima2022discovery} and most recently, by nuclear magnetic resonance (NMR) studies\cite{Yu2023}.

Possible origins of such residual DOS in the superconducting states of the heavily S-substituted FeSe has been discussed in the aforementioned Ref.~\cite{nagashima2022discovery,Yu2023}.
Impurity effects or coexistence of spatially separated SC and normal phase are excluded, because the samples are clean and homogeneous as seen from  quantum oscillation\cite{Coldea2019} and STM experiments\cite{Hanaguri2018}. For measurements done under external field such as the NMR measurements, another possible explanation for the observed residual DOS is the Volovik effect\cite{Volovik1993,Bang2012}. However, the Volovik effect cannot account for the order of magnitude difference in the relaxation rate across samples with different substitution levels but the same external field. It has been suggested\cite{Setty2019,Setty2020,Cao2023} that the so-called ultranodal superconducting state, which by definition hosts Bogoliubov Fermi surfaces (BFS), is responsible for the large residual DOS in these systems. 

Ultranodal states are superconducting states with extended gap nodes that, in contrast to usual point nodes or line nodes in three dimensions, have the same dimension as the underlying normal state Fermi surface. Such extended nodes are called Bogoliubov Fermi surfaces\cite{VOLOVIK1989,Timm2017, Brydon2017, Brydon2018,Sumita2019,Autti2020}. The existence of BFS does not necessarily require non-trivial topology, as is the case in Ref.~\cite{wu_private}, but they are topologically protected by a $\mathbb{Z}_2$ invariant if the superconducting state possesses inversion symmetry. In a multiband spin-$1/2$ superconductor, BFS can arise from an interband non-unitary triplet pairing term or from a magnetic order that breaks time-reversal symmetry, and may\cite{Setty2019} or may not\cite{Cao2023,wu_private} preserve the inversion symmetry. It has also been shown\cite{Cao2023} that the non-unitarity of the interband triplet pairing can be induced by driving the system close to a magnetic instability, in which case the magnetic moment of the the non-unitary triplet pair aligns with the fluctuating magnetic order.

The existence of a BFS explains well the residual DOS in the tetragonal Fe(Se,S) as seen from specific heat or STS experiments (see however Ref.~\cite{Islam2023} for an alternative picture), as well as the possible $C_4$ symmetry breaking in the superconducting phase as seen in the ARPES experiment\cite{nagashima2022discovery}. However, it has not been fully understood how the ultranodal scenario can fit the recent NMR data presented in Ref.~\cite{Yu2023}.

The NMR measurement in Ref.~\cite{Yu2023}, performed on FeSe$_{1-x}$S$_x$ at several S-substitution levels across the nematic QCP with in-plane applied field and temperature down to $100$\;mK, shows not only a finite value of $1/(T_1T)$ at zero temperature for the $x=0.18$ and $x=0.23$ samples, but also an unusual upturn as temperature decreases towards zero. While the former can be understood fairly straightforwardly as yet another signature of the zero energy residual DOS in these materials, the latter requires a more sophisticated understanding. In this paper, we study the models for BFS systems discussed in \cite{Setty2019,Setty2020,Cao2023} to further calculate the spin fluctuations in the ultranodal states. We compare our calculations of $1/(T_1T)$ to the experimental data, and show that the upturn is likely due to the interplay between strong magnetic fluctuation and multiband physics in such systems.

\begin{figure*}[!ht]
    \centering
    \includegraphics[width=\textwidth]{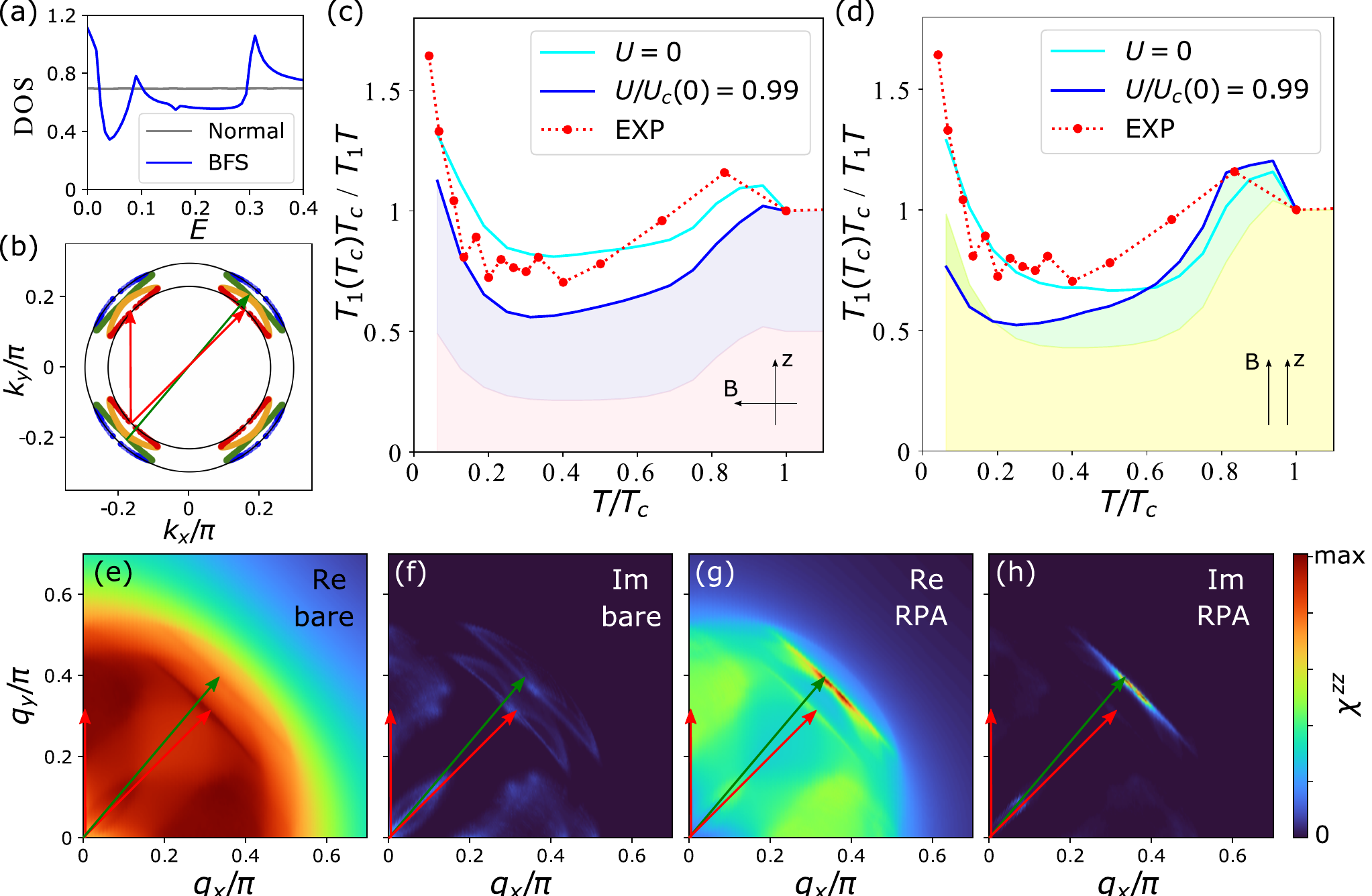}
    \caption{(a) Zero temperature DOS for the two-band model \eqref{Hmf}, with $\mu_1=3.5$, $\mu_2=3.2$ and $\Delta_1(\mathbf k)=\Delta_2(\mathbf k)=0.06|\cos 2\theta_\mathbf{k}|$, where $\theta_{\mathbf{k}}$ is the angle of $\mathbf{k}$ on the 2D Fermi surface. $\Delta_{\uparrow\uparrow}(\mathbf k)=0.15$, $\Delta_{\downarrow\downarrow}(\mathbf k)=0$ at $T=0$. (b) Normal (black) and Bogoliubov (colored) Fermi surfaces. Within the same color of the BFS the coherent scattering amplitudes are larger than between points on BFSs with different color (See main text Sec. III).  Arrows shows some of the dominant scattering processes as seen from panel (e-h). (c) Normalized $1/(T_1T)$ below $T_c$ assuming $\hat{z}$ perpendicular to the applied magnetic field. Cyan and blue curves calculated from bare and RPA susceptibility near a magnetic instability respectively. Red dots are experimental data taken from Ref.~\cite{Yu2023}. The pink and blue shaded area represent contributions from $\mathop{\text{Im}}\chi^{yy}$ and $\mathop{\text{Im}}\chi^{zz}$. $T_c$ is taken to be $0.08$. The critical $U$ is determined from the normal and Bogoliubov band structure, and $U_c(T_c)=10.8$, $U_c(0)=9.6$. (d) Normalized $1/(T_1T)$ below $T_c$ assuming $\hat{z}$ parallel to the applied magnetic field. The yellow and green shaded area represent contributions from $\mathop{\text{Im}}(\chi^{xx}+\chi^{yy})$ and $\mathop{\text{Re}}\chi^{xy}$. Note that the contribution from $\mathop{\text{Re}}\chi^{xy}$ at low $T$ is negative. (e-h) Real and imaginary parts of the spin susceptibility at $T=0$. The arrows are the same as in panel (a). One can see that there is a shift of the dominant contribution from the red arrow to green arrow as $U$ increases. The color bar maxima are 0.7, 0.03, 8.5, 9 for panel (e), (f), (g), (h) respectively.}
    \label{Fig1}
\end{figure*}

\section{Model}
We adopt a minimal two-band mean field model with intraband spin-singlet pairing and interband non-unitary spin-triplet pairing from previous works\cite{Setty2019,Setty2020,Cao2023}
\begin{align*}
    H=&\sum_{\mathbf{k},\sigma,i}\epsilon_{i\mathbf{k}\sigma}c^\dagger_{i\mathbf{k}\sigma}c_{i\mathbf{k}\sigma}-\sum_{\mathbf{k},i}\Delta_{i}(\mathbf{k})(c^\dagger_{i\mathbf{k}\uparrow}c^\dagger_{i-\mathbf{k}\downarrow}+h.c.)\\
    &-\sum_{\mathbf{k},\sigma}\Delta_{\sigma\sigma}(\mathbf{k})(c^\dagger_{1\mathbf{k}\sigma}c^\dagger_{2-\mathbf{k}\sigma}+h.c.).
    \numberthis \label{Hmf}
\end{align*}
Here $i=1,2$ is the band index.
Seeking  qualitative results at low temperatures, we make the assumption that all gaps correspond to a single $T_c$ and follow a BCS-like temperature dependence, where the key feature is that the deviation from the $T=0$ value is exponentially or power law small at low temperature.
Also, for simplicity, we consider a tight-binding model with only nearest neighbor hopping and $\epsilon_{i\mathbf{k}\sigma}=2(\cos k_x+\cos k_y)-\mu_i$. We have set the nearest neighbor hopping parameter $t=1$ and adopt it as our unit of energy throughout the calculations below.

We calculate the spin susceptibility
\begin{align}
    \chi^{uv}(\mathbf{q},t)&=i \theta(t)\sum_{\mathbf{q'}}\langle\left[S^u(\mathbf{q},t), S^v(\mathbf{q'},0)\right]\rangle,
\end{align}
where $S^u(\mathbf{q},t=0)=\sum_{i,\mathbf{k},\alpha,\beta}c^\dagger_{i\mathbf{k}\alpha}\sigma^u_{\alpha\beta}c_{i\mathbf{k+q}\beta}$ is the total spin operator summed over the two bands, and $u,v=x,y,z$. 
The spin quantization axis ($z$ axis of the Pauli matrices) denotes the direction of the magnetic moment of our non-unitary triplet pair\cite{Cao2023},   
which breaks the spin rotational symmetry of the ultranodal state and makes the $z$-direction inequivalent to the $x,y$-directions. Depending on the angle between the $z$-direction and the applied magnetic field, all components of $\chi^{uv}$ might contribute to the longitudinal relaxation time $T_1$\cite{Moriya1956}. We focus on two configurations: The external field $\vec{B}$ is either parallel to or perpendicular to the $\hat{z}$-direction when calculating $T_1$, and all the other configurations should give results in between these two configurations. For $\vec{B}\parallel\hat{z}$,
\begin{align}
    \frac{1}{T_1}\propto T\lim_{\omega\rightarrow0}\sum_\mathbf{q}\frac{\operatorname{Im}\chi^{+-}(\mathbf{q},\omega)}{\omega}
    \label{1/t1},
\end{align}
where $+/-$ denotes $x\pm iy$. Similarly for $\vec{B}\parallel\hat{x}$ Eq. \eqref{1/t1} is still valid except $+/-$ now denotes $y\pm iz$.

To this end, we first find the Nambu Green's function $G_\mathbf{k}(\omega)$ by diagonalizing the Nambu Hamiltonian corresponding to Eq.~\eqref{Hmf}.
The Nambu basis we use is $\psi_\mathbf{k}=[c_{1\mathbf{k}\uparrow},  c_{1\mathbf{-k}\downarrow}, c_{2\mathbf{-k}\uparrow}, c_{2\mathbf{k}\downarrow},c_{1\mathbf{k}\uparrow}^\dagger, c_{1\mathbf{-k}\downarrow}^\dagger,c_{2\mathbf{-k}\uparrow}^\dagger, c_{2\mathbf{k}\downarrow}^\dagger]^T$.  
With the eigenvalues $E_{l\mathbf{k}}$ and the eigenvector matrix $U_\mathbf{k}$ of the Nambu Hamiltonian, the Nambu Green's function can be expressed as
\begin{align}
    G_\mathbf{k}(\omega)=U_\mathbf{k}^\dagger\text{diag}\left(\frac{1}{\omega-E_{1\mathbf{k}}},\  \ldots,\ \frac{1}{\omega-E_{8\mathbf{k}}}\right)U_\mathbf{k}.
    \label{Green_fctn}
\end{align}
At this point, we would like to also define the $8\times8$ Nambu spin matrices $\Sigma^u\equiv \text{diag}(\sigma^u,\sigma^u,-(\sigma^u)^T,-(\sigma^u)^T)$, composed of $2\times2$ Pauli matrices on their diagonal blocks. The bare spin-spin correlation function in the Matsubara representation is $C^{uv}(\mathbf{q},i\nu_m)=-\frac{1}{2}\frac{1}{\beta}\sum_{i\omega_n} \sum_{\mathbf{k}}\operatorname{Tr}\left(\Sigma^uG_\mathbf{k+q}(i\omega_n+i\nu_m)\Sigma^vG_\mathbf{k}(i\omega_n)\right)$, where $\nu_m$ is a bosonic frequency. Substituting in Eq.\eqref{Green_fctn} and performing the Matsubara sum and the analytic continuation to the real axis, we obtain first the bare density-density bubble in the quasiparticle band space
\begin{align*}
    \chi_{l l'm m'}^{(0)}(\mathbf{q},\omega)=-\sum_{\mathbf{k},r,s}U_{rl\mathbf{k}}U_{s l'\mathbf{k+q}}^*U_{rm\mathbf{k}}^*U_{s m'\mathbf{k+q}}\\
    \times\frac{f(E_{s\mathbf{k+q}})-f(E_{r\mathbf{k}})}{E_{s\mathbf{k+q}}-E_{r\mathbf{k}}-\omega-i0^+}
    \numberthis\label{bareChi}.
\end{align*}
Then the $zz$-component of the spin susceptibility can be written as
\begin{align}
    \chi^{(0)zz}(\mathbf{q},\omega)=\frac{1}{2}\sum_{l,m}\Sigma^z_{ll}\Sigma^z_{mm}\chi_{llmm}^{(0)}(\mathbf{q},\omega),
    \label{bareSpinSus}
\end{align}
and for the other components $u,v=x,y$ we have
\begin{align}
    \chi^{(0)uv}(\mathbf{q},\omega)=\frac{1}{2}\sum_{l,m}\Sigma^u_{l\bar l}\Sigma^v_{\bar mm}\chi_{l\bar lm\bar m}^{(0)}(\mathbf{q},\omega),
    \label{bareSpinSus2}
\end{align}
where $\bar{l}$ denotes the Nambu index that corresponds to the time-reversed $l$th operator in the Nambu basis. For example, $\bar{1}=2$ and $\bar{8}=7$. From Eq. \eqref{bareSpinSus} and \eqref{bareSpinSus2} we see that two types of $\chi_{l l'm m'}^{(0)}$ are particularly important, namely the $llmm$-type and $l\bar lm\bar m$-type. Accordingly let us define two types of coherence factors
\begin{align}
    W_{ll}(r\mathbf{k},s\mathbf{k'})\equiv U_{rl\mathbf{k}}U_{sl\mathbf{k'}}^*\label{cf}\\
    W_{l\bar l}(r\mathbf{k},s\mathbf{k'})\equiv U_{rl\mathbf{k}}U_{s\bar l\mathbf{k'}}^*\label{cf2}
\end{align}
where $r\mathbf{k}$ is a composite label referring to the Bogoliubov quasiparticle at momentum $\mathbf{k}$ in the $r$th quasiparticle band. They will be useful when we analyze the structure of the Bogoliubov Fermi surfaces later.

We can further investigate using a random phase approximation (RPA) the effect of a residual interaction in the particle-hole channel. We consider an interaction of the Hubbard type
\begin{align*}
H_U&=\frac{1}{2}\sum_{\mathbf{r},i}Un_{i\mathbf{r}\uparrow}n_{i\mathbf{r}\downarrow}\numberthis\label{Hubbard}
\\
&=\frac{1}{16}\sum_\mathbf{k,k',q}\sum_{l,m,l',m'}\Gamma_{ll'mm'}\psi^\dagger_{l\mathbf{k}}\psi_{l'\mathbf{k-q}}\psi^\dagger_{m'\mathbf{k'}}\psi_{m\mathbf{k'+q}}
\end{align*}
In the last step we have rewritten the interaction using the Nambu basis and defined a coupling tensor $\Gamma$  that does not depend on the momentum transfer $\mathbf{q}$.
The non-zero elements of $\Gamma$ are $\Gamma_{ll\bar l\bar l}=U$ and $\Gamma_{l\bar ll\bar l}=-U$ for $l=1,2,..,8$.
As shown in \cite{supplement}, it is sufficient to consider only the $8\times8$ $llmm$ matrix blocks (with $ll$ being the row index and $mm$ being the column index) of $\chi^{(0)}_{ll'mm'}$ and $\Gamma_{ll'mm'}$, which I denote as $\hat\chi^{(0)}$ and $\hat\Gamma$, for calculating $\chi^{\mathrm{(RPA)}zz}$. For $\chi^{\mathrm{(RPA)}uv}$ with $u,v=x,y$, it is sufficient to consider only the $8\times8$ $l\bar lm\bar m$ blocks of $\chi^{(0)}_{ll'mm'}$ and $\Gamma_{ll'mm'}$, which I denote as $\tilde\chi^{(0)}$ and $\tilde\Gamma$.
The RPA density-density bubble is related to the bare bubble through
\begin{align}
    \hat{\chi}^{\mathrm{(RPA)}}(\mathbf{q},\omega)=\hat{\chi}^{(0)}(\mathbf{q},\omega)\left(\hat{I}+\hat{\Gamma}\hat{\chi}^{(0)}(\mathbf{q},\omega)\right)^{-1};\\
    \tilde{\chi}^{\mathrm{(RPA)}}(\mathbf{q},\omega)=\tilde{\chi}^{(0)}(\mathbf{q},\omega)\left(\tilde{I}+\tilde{\Gamma}\tilde{\chi}^{(0)}(\mathbf{q},\omega)\right)^{-1}.
    \label{chiRPA}
\end{align}
The sign convention for the above equation is also explained in detail in \cite{supplement}. Here we note only that the usual RPA sign emerges in the more standard spin basis.
The RPA spin susceptibility is 
\begin{align*}
    \chi^{\mathrm{(RPA)}zz}(\mathbf{q},\omega)=\frac{1}{2}\sum_{l,m}\Sigma^z_{ll}\Sigma^z_{mm}\hat{\chi}_{llmm}^{\mathrm{(RPA)}}(\mathbf{q},\omega);\numberthis\label{RPAspinSus}\\
    \chi^{\mathrm{(RPA)}uv}(\mathbf{q},\omega)=\frac{1}{2}\sum_{l,m}\Sigma^u_{l\bar l}\Sigma^v_{\bar mm}\tilde{\chi}_{l\bar lm\bar m}^{\mathrm{(RPA)}}(\mathbf{q},\omega),\\
    u,v=x,y
    \numberthis\label{RPAspinSus2}
\end{align*}
by analogy to Eq. \eqref{bareSpinSus} and \eqref{bareSpinSus2} with $\chi^{(0)}\rightarrow \hat\chi^{\mathrm{(RPA)}}$ or $\tilde\chi^{\mathrm{(RPA)}}$.

\section{Results}

We numerically calculated the spin susceptibility of the ultranodal states for the model Hamiltonian. To summarize the result, we found that the bare susceptibility calculation always give rise to non-zero residual $1/(T_1T)$ at zero temperature when BFSs are present, as expected due to the zero energy residual DOS. However, the bare $1/(T_1T)$ rarely increases as temperature decreases near $T=0$, unless van Hove singularities of the Bogoliubov quasiparticle bands are tuned to the Fermi level, contributing to a large zero energy peak in the DOS. 
On the other hand, if we take into account the  correlation effects using the RPA calculation, certain scattering between  coherent spots/segments on the BFS can get strongly enhanced, resulting in an upturn in the $1/(T_1T)$ as temperature decreases, even when the zero energy DOS is not peaked, or when the BFS are not strongly nested. Below we discuss in details these results.

\begin{figure*}[!ht]
    \centering
    \includegraphics[width=\textwidth]{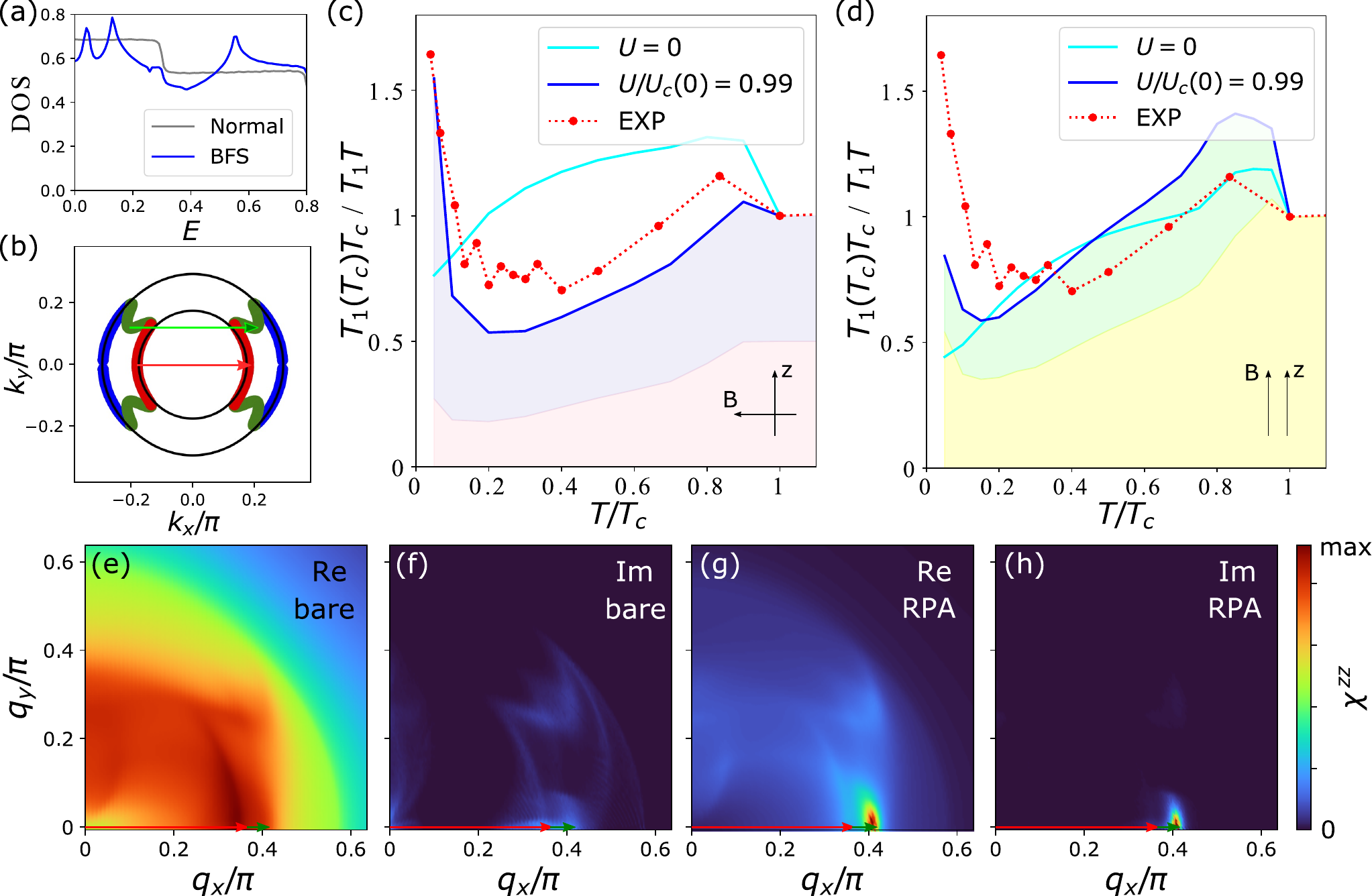}
    \caption{Same as Fig. \ref{Fig1} but with a different set of parameters: $\mu_1=3.7$, $\mu_2=3.2$, $\Delta_1(\mathbf k)=\Delta_2(\mathbf k)=0.05$, $\Delta_{\uparrow\uparrow}(\mathbf k)=0.3\cos \theta_\mathbf{k}$, $\Delta_{\downarrow\downarrow}(\mathbf k)=0$, $T_c=0.1$. $U_c(T_c)=10.8$, $U_c(0)=10.1$. The color bar maxima are 0.7, 0.026, 35, 11 for panel (e), (f), (g), (h) respectively.}
    \label{Fig2}
\end{figure*}

In Figs. \ref{Fig1} and \ref{Fig2} we show in parallel two examples of having upturns in the $1/(T_1T)$ at low temperature as a result of correlations, the existence of BFS and multiband effects combined. Fig. \ref{Fig1} corresponds to a scenario where the intraband singlet
$\Delta_{i}(\mathbf{k})$ is taken to be nodal s-wave with accidental nodes along the 45 degree directions, and the interband triplet pairing $\Delta_{\uparrow\uparrow}(\mathbf{k})$ isotropic. Fig. \ref{Fig2} corresponds 
to the $C_2$ symmetric scenario discussed in Ref.~\cite{Cao2023}, where the interband triplet pairing $\Delta_{\uparrow\uparrow}(\mathbf{k})$ is assumed to be p-wave and the intraband singlet pairing $\Delta_{i}(\mathbf{k})$ is taken to be isotropic for simplicity. A BFS then forms only when $\Delta_{\up\up}$ is sufficiently large.

In both cases we have set $\Delta_{\downarrow\downarrow}(\mathbf{k})=0$. In Fig. \ref{Fig1}, the bare susceptibility already gives rise to upturns in the $1/(T_1T)$ (panel (c) and (d) cyan curve,  each corresponds to different orientations of the external field). This is because the van Hove singularity (band extremum corresponding to where the superconducting gap opens) of the Bogoliubov band has been tuned at the Fermi level by changing the interband order parameter, and a peak in the quasiparticle density of states exists at exactly zero energy (Fig. \ref{Fig1} panel (a)). We found that upturns in the bare $1/(T_1T)$ at low temperature seem to be always associated with such peaks at zero energy in the DOS. Although such peaks in the DOS are not desired as they are fine-tuned and not consistent with the spectroscopic data\cite{Hanaguri2018}, they are not required once we include  correlations. This can be seen in Fig. \ref{Fig2} (a) through (d), where  the density of states is not peaked at zero energy and the bare $1/(T_1T)$ curve does not have an upturn while the RPA curve close to magnetic instability does. The qualitative difference between the $U=0$ curve and the $U \lessapprox U_c(0)$ curve in Fig. \ref{Fig2} (c) and (d) is unusual, since normally one would expect, from its simplest form for the normal metal as in Eq. (S.22)\cite{supplement}, the RPA susceptibility to be enhanced further where the bare susceptibility is already large.

To better understand this unusual behavior,
we analyze here the major contribution to the upturns in Fig. \ref{Fig1} and \ref{Fig2} panel (c) from $\chi^{zz}$ (blue shaded area), and provide in \cite{supplement} the same analysis for the contributions from $\chi^{yy}=\chi^{xx}$ (pink or yellow shaded area in panel (c) or (d)). We first divide the BFS into several segments, as shown by the color scheme in panel (b) of Fig. \ref{Fig1} and \ref{Fig2}, within each color the scattering that preserves the spin and band is much stronger than across different colors. This is done by treating the $\mathbf{k}$-points on the BFS as vertices of an weighted undirected graphs with weights given by linear combinations of the eight $ll$-type coherence factors $W_{ll}$ defined in Eq. \eqref{cf}, and employing the Leiden algorithm for community detection\cite{traag2019louvain,csardi2006igraph}. We see that the parts of the BFS that follow the shape of the  normal Fermi surfaces (red and blue in Fig. \ref{Fig1} and \ref{Fig2} panel (b)) are well separated from the rest of the BFSs in terms of non-spin-flip scattering processes. Then in Fig. \ref{Fig1} and \ref{Fig2} panel (e,f) we plot the  $zz$-component of the bare spin susceptibility as in Eq. \eqref{bareSpinSus} at zero temperature, and in panel (g,h) we show the RPA spin susceptibility \eqref{RPAspinSus} at $U \lessapprox U_c(0)$ and zero temperature. We identify the important $\mathbf{q}$ vectors as the red and green arrows connecting segments of BFSs with the same color shown in Fig. \ref{Fig1} and \ref{Fig2} panel (b). From  panel (e) we first see that the real part of the bare susceptibility is the largest at the $\mathbf{q}$ vectors connecting the red part of the BFSs, but is not strongly peaked at any particular $\mathbf{q}$ vector. The latter observation is an indication of no strong nesting between the the BFSs. Secondly, by comparing panel (e,f) with panel (g,h) in Fig. \ref{Fig1} and \ref{Fig2}, we see that although the $\mathbf{q}$ vectors that connect the green part of the BFSs are only subdominant in the bare susceptibility, they become the dominant $\mathbf{q}$ vectors near the magnetic instability. This shift of the dominant $\mathbf{q}$ vectors as the interaction $U$ increases within an RPA calculation can only be explained by nontrivial multiband effects embedded in the coherence factor $W_{ll}$, which is consistent with the unusual change in the shape of the normalized $1/(T_1T)$ curve as $U$ increases. 

In case the external field fully polarizes the magnetic moment of the non-unitary triplet pair so that $\vec{B}\parallel\hat{z}$, $\chi^{zz}$ will not be responsible for the longitudinal relaxation at all. Nevertheless, as seen from panel (d) in Fig. \ref{Fig1} and \ref{Fig2}, there will still be a minor upturn in $1/(T_1T)$ due to contributions from $\chi^{xx}=\chi^{yy}$. They share some key features with $\chi^{zz}$ as discussed in the previous paragraph\cite{supplement}: First, there can be shifts of dominant $\mathbf{q}$ vector as $U$ increases. Second,
only those $\mathbf{q}$ vectors connecting coherent parts of the BFS, which usually carry interband character,  contribute significantly to the susceptibility at $T=0$. Therefore, the low temperature upturns in panel (d) of Fig. \ref{Fig1} and \ref{Fig2} have the same origin as in panel (c), namely it is due to the existence of BFS, correlation and multiband physics.

\section{Conclusion}

To summarize, we have calculated the spin susceptibilities for the ultranodal states in a minimal two-band model, where the interband non-unitary spin-triplet pairing is responsible for the Bogoliubov Fermi surfaces.  We found that the existence of BFSs in such models naturally gives rise to finite residual value in the $1/(T_1T)$ at zero temperature, but does not necessarily produce the large upturns at low temperature, as seen in the experiments \cite{Yu2023} on the Fe(Se,S) system, in a non-interacting  calculation. We then studied the effect of correlation within random phase approximation in the ultranodal state. By adding a Hubbard interaction in the particle-hole channel while not changing the pre-assumed pairing gaps, we see that the spin susceptibilities at $\mathbf{q}$ vectors connecting coherent segments/spots on the BFS get strongly enhanced at low temperature when the interaction is strong, resulting in upturns in $1/(T_1T)$ irrespective of the presence or absence of upturns in the bare calculation. These spots have strong interband character as indicated from their position on the BFSs, and do not have particularly large contribution to the spin susceptibilities at weak interaction.
The $1/(T_1T)$ calculated from the spin susceptibilities close to the antiferromagnetic instability shows robust upturns at low temperature for all orientations of external field. Although the upturn is the smallest when the Cooper pair moment aligns with the external field, we expect that spin orbit coupling or strong antiferromagnetic fluctuations could drive the system away from the perfectly polarized configuration. Therefore, we conclude that the experimentally observed upturn in $1/(T_1T)$ can be explained as a combined effect of the presence of BFS, interband physics and correlation.

Our theory is primarily applicable to the tetragonal phase of FeSe$_{1-x}$S$_x$ with $x>0.17$ at ambient pressure. For the nematic phase with $x<0.17$ at ambient pressure and the tetragonal phase with $x<0.17$ under pressure\cite{Rana2020}, the low temperature $1/(T_1T)$ seems to have a Korringa behavior, i.e. constant in temperature,  with smaller but finite residual values. Our calculation of the $1/(T_1T)$ is consistent with these data assuming weak correlation or small BFSs, but whether the ultranodal scenario can apply to these situations requires a more careful and comprehensive study in the future.

\section{Acknowledgements}  L.~F. acknowledges support by the European Union's Horizon 2020 research and innovation programme through the Marie Sk\l{}odowska-Curie grant SuperCoop (Grant No 838526). P.J.H. and Y.C. were  supported by  DOE grant number DE-FG02-05ER46236.  A.K. acknowledges support by the Danish National Committee for Research Infrastructure (NUFI) through the ESS-Lighthouse Q-MAT.\\ \newline
\noindent

\bibliography{NMR.bib}
\nocite{Romer2019,Romer2022}

\end{document}


\title{Supplementary Material: Spin fluctuations in the ultranodal superconducting state of Fe(Se,S)}
\author{Yifu Cao$^1$, Chandan Setty$^2$, Andreas Kreisel$^3$, Laura Fanfarillo$^{1,4}$ and P.J. Hirschfeld$^1$}
\affiliation{$^1$Department of Physics, University of Florida, Gainesville, Florida 32603, USA}

\affiliation{$^2$Department of Physics and Astronomy, Rice Center for Quantum Materials, Rice University, Houston, Texas 77005, USA}

\affiliation{$^3$Niels Bohr Institute, University of Copenhagen, Jagtvej 155 , DK-2200, Copenhagen, Denmark}

\affiliation{$^4$Istituto dei Sistemi Complessi (ISC-CNR), Via dei Taurini 19, I-00185 Rome, Italy}

\maketitle

\setcounter{equation}{0}
\renewcommand{\theequation}{S.\arabic{equation}}
\renewcommand{\thefigure}{S\arabic{figure}}
In the supplementary material we provide information about how the standard multiorbital RPA formalism in e.g. Ref.~\cite{Romer2019,Romer2022} gives rise to the $8\times8$ matrix formalism in Nambu space we used in the main text, and about the $\chi^{xx}=\chi^{yy}$ components of the spin susceptibility.

\section{RPA formalism}
We start by considering a bare particle-hole bubble in the general form
\begin{align*}
    \chi_{ll'mm'}^{(0)}(\mathbf{q},\omega)=-\sum_{\mathbf{k},r,s}U_{rl\mathbf{k}}U_{sl'\mathbf{k+q}}^*U_{rm\mathbf{k}}^*U_{sm'\mathbf{k+q}}\frac{f(E_{s\mathbf{k+q}})-f(E_{r\mathbf{k}})}{E_{s\mathbf{k+q}}-E_{r\mathbf{k}}-\omega-i0^+}\,,
    \numberthis\label{A_bareChi}
\end{align*}
where $E_{l\mathbf{k}}$ are the eigenvalues, $U_\mathbf{k}$ is the eigenvector matrix of the Nambu Hamiltonian and $f(x)=1/[\exp(\beta x))+1]$ is the Fermi function.
$\chi_{ll'mm'}^{(0)}(\mathbf{q},\omega)$ transforms like a $8\times8\times8\times8$ rank-4 tensor as the Nambu basis transforms, while contracts like a $64\times64$ rank-2 tensor (matrix with $ll'$ a composite row index and $mm'$ a composite column index) in the RPA series. The Hubbard interaction we are considering is
\begin{align}
H_U=\frac{1}{2}\sum_{\mathbf{r},i}Un_{i\mathbf{r}\uparrow}n_{i\mathbf{r}\downarrow}
=\frac{1}{16}\sum_{\mathbf{k,k',q},l,m,l',m'}\Gamma_{ll'mm'}\psi^\dagger_{l\mathbf{k}}\psi_{l'\mathbf{k-q}}\psi^\dagger_{m'\mathbf{k'}}\psi_{m\mathbf{k'+q}}
\label{A_Hubbard}
\end{align}
where on the RHS the terms quadratic in $c$,$c^\dagger$ has been discarded since they only renormalize the chemical potential. Let's represent $\Gamma_{ll'mm'}$ as a $64\times64$ matrix which only has two $8\times8$ non-zero blocks, ordered as following:
\begin{align}
    \breve \Gamma\equiv[\Gamma_{ll'mm'}]_{64\times64}=\begin{pmatrix}
        [\Gamma_{llmm}]_{8\times8}&0_{8\times48}&0_{8\times8}\\
        0_{48\times8}&0_{48\times48}&0_{48\times8}\\
        0_{8\times8}&0_{8\times48}&[\Gamma_{l\bar{l}m\bar{m}}]_{8\times8}
        \label{A_gamma}
    \end{pmatrix}
\end{align}
We define $\bar{l}$ as the Nambu index that corresponds to the time-reversed $l$th operator in the Nambu basis. For example, $\bar{1}=2$ and $\bar{8}=7$. $[\Gamma_{llmm}]_{8\times8}$ is a block diagonal matrix, $[\Gamma_{llmm}]_{8\times8}=U\text{diag}[\sigma^x,\sigma^x,\sigma^x,\sigma^x]$, and $[\Gamma_{l\bar{l}m\bar{m}}]_{8\times8}=-UI$. We use hat and tilde to denote these two kinds of $8\times8$ blocks so that $\hat{\Gamma}\equiv[\Gamma_{llmm}]_{8\times8}$ and $\Tilde{\Gamma}\equiv[\Gamma_{l\bar{l}m\bar{m}}]_{8\times8}$. Similarly $\chi_{ll'mm'}^{(0)}(\mathbf{q},\omega)$ can be represented as $64\times64$ matrices which we denote as $\breve\chi^{(0)}(\mathbf{q},\omega)$. The RPA spin susceptibility is then
\begin{align}
    \chi^{\mathrm{(RPA)}uv}(\mathbf{q},\omega)=\frac{1}{2}{S^u}\Breve\chi^{(0)}(\mathbf{q},\omega)\left(\Breve{I}+\Breve\Gamma\Breve\chi^{(0)}(\mathbf{q},\omega)\right)^{-1}{S^v}^\dagger
    \label{A_chiRPAss}
\end{align}
The spin vertices have the following forms accordingly:
\begin{align}
     S^x=(0_{1\times8},\quad0_{1\times48},\quad D^x)\label{5-S_begin}\\
     S^y=(0_{1\times8},\quad0_{1\times48},\quad D^y)\\
     S^z=(D^z,\quad0_{1\times48},\quad0_{1\times8})\\
    D^x=(1,1,1,1,-1,-1,-1,-1)\\
    D^y=(-i,i,-i,i,-i,i,-i,i)\\
    D^z=(1,-1,1,-1,-1,1,-1,1) \label{A_S_end}
\end{align}
where the vectors $D^u$ are related to the matrices $\Sigma^u$ defined in the main text by $D^z_l=\Sigma^z_{ll}$ and $D^v_l=\Sigma^v_{l\bar l}$ for $v=x,y$. For the bare susceptibility $\Breve\chi^{(0)}$, we note that for any $l,m$, $\chi^{(0)}_{llm\bar m}=\chi^{(0)}_{l\bar lmm}=0$. In terms of the $3\times3$ blocks this means
\begin{align}
    \Breve\chi^{(0)}=\begin{pmatrix}
        [\chi^{(0)}_{llmm}]_{8\times8}&*&0_{8\times8}\\
        *&*&*\\
        0_{8\times8}&*&[\chi^{(0)}_{l\bar{l}m\bar{m}}]_{8\times8}
    \end{pmatrix}
    \label{A_chi0}
\end{align}
Here $*$‘s are placeholders for unimportant blocks. Eq. \eqref{A_chi0} is true because the pair of Green's functions $G_{\bar{m}l\mathbf{k}}(\omega')$ and $G_{lm\mathbf{k+q}}(\omega'+\omega)$ cannot be both non-zero at the same time, given that there is only the intraband singlet pairing and interband triplet pairing processes in the Hamlitonian (Eq.~(1) in the main text). Again we denote $\hat{\chi}^{(0)}\equiv[{\chi}^{(0)}_{llmm}]_{8\times8}$ and $\Tilde{\chi}^{(0)}\equiv[{\chi}^{(0)}_{l\bar{l}m\bar{m}}]_{8\times8}$. By directly multiplying the matrices, one can verify that
\begin{align}
    \Breve\chi^{(0)}\Breve\Gamma\Breve\chi^{(0)}=\begin{pmatrix}
        \hat\chi^{(0)}\hat\Gamma\hat\chi^{(0)}&*&0_{8\times8}\\
        *&*&*\\
        0_{8\times8}&*&\Tilde\chi^{(0)}\Tilde\Gamma\Tilde\chi^{(0)}
    \end{pmatrix}\label{5-ChiGammaChi}
\end{align}
From Eq. \eqref{5-ChiGammaChi} one can inductively deduce that the two diagonal $8\times8$ blocks in the RPA susceptibility $\Breve{\chi}^\mathrm{(RPA)}$ can be separately calculated using only the knowledge of the diagonal $8\times8$ blocks in the bare susceptibility $\Breve{\chi}^{(0)}$. To be more specific,
\begin{align}
    \Breve\chi^{\mathrm{(RPA)}}(\mathbf{q},\omega)&\equiv\Breve\chi^{(0)}(\mathbf{q},\omega)\left(\Breve{I}+\Breve\Gamma\Breve\chi^{(0)}(\mathbf{q},\omega)\right)^{-1}\\
    &=\begin{pmatrix}
        \hat\chi^{\mathrm{(RPA)}}(\mathbf{q},\omega)&*&0_{8\times8}\\
        *&*&*\\
        0_{8\times8}&*&\Tilde\chi^{\mathrm{(RPA)}}(\mathbf{q},\omega)\label{5-chiRPA}
    \end{pmatrix}
\end{align}
where $\hat\chi^{\mathrm{(RPA)}}(\mathbf{q},\omega)=\hat\chi^{(0)}(\mathbf{q},\omega)\left(\hat{I}+\hat\Gamma\hat\chi^{(0)}(\mathbf{q},\omega)\right)^{-1}$ and the same for the tilde blocks. It is now also easy to see from Eq.~\eqref{A_chiRPAss} that for $u,v=x,y$, the RPA spin susceptibility only involves the tilde $8\times8$ blocks;
\begin{align}
    \chi^{\mathrm{(RPA)}uv}(\mathbf{q},\omega)=\frac{1}{2}D^u\Tilde\chi^{(0)}(\mathbf{q},\omega)\left(\Tilde{I}+\Tilde\Gamma\Tilde\chi^{(0)}(\mathbf{q},\omega)\right)^{-1}{D^v}^\dagger,\quad\quad u,v=x,y\label{5-chixy}
\end{align}
for $u,v=z$,  $\chi^{\mathrm{(RPA)}zz}(\mathbf{q},\omega)$ involves only the hatted $8\times8$ blocks; 
\begin{align}
    \chi^{\mathrm{(RPA)}zz}(\mathbf{q},\omega)=\frac{1}{2}D^z\hat\chi^{(0)}(\mathbf{q},\omega)\left(\hat{I}+\hat\Gamma\hat\chi^{(0)}(\mathbf{q},\omega)\right)^{-1}{D^z}^\dagger\label{5-chizz}
\end{align}
and $\chi^{\mathrm{(RPA)}zx}(\mathbf{q},\omega)=\chi^{\mathrm{(RPA)}zy}(\mathbf{q},\omega)=0$.

As a further sanity check, let's take a one band model and turn off the pairing amplitude. The Nambu basis reduces to $[c_{\mathbf{k}\uparrow},c_{\mathbf{k}\downarrow},c^\dagger_{\mathbf{-k}\uparrow},c^\dagger_{\mathbf{-k}\downarrow}]$. The $16\times16$ bare susceptibility matrix contains two nonzero blocks: the hatted block $\hat{\chi}^{(0)}\equiv[{\chi}^{(0)}_{llmm}]_{4\times4}$ and the tilde block $\Tilde{\chi}^{(0)}\equiv[{\chi}^{(0)}_{l\bar{l}m\bar{m}}]_{4\times4}$, both proportional to the identity matrix
\begin{align}
    \hat{\chi}^{(0)}=\Tilde{\chi}^{(0)}=\chi^{(0)}I
\end{align}
Since both the $\hat{\chi}^{(0)}$ and $\hat{\Gamma}$ (and also both $\tilde{\chi}^{(0)}$ and $\tilde{\Gamma}$) are block diagonal in particle-hole space, with the two diagonal blocks being the same, Eq. \eqref{5-chixy} and \eqref{5-chizz} reduce to $2\times2$ matrix equations in the spin space
\begin{align}
    \chi^{\mathrm{(RPA)}xx}(\mathbf{q},\omega)=\begin{pmatrix}
        1&1
    \end{pmatrix}\chi^{(0)}(\mathbf{q},\omega)\begin{pmatrix}
        1-U\chi^{(0)}(\mathbf{q},\omega)&0\\
        0&1-U\chi^{(0)}(\mathbf{q},\omega)
    \end{pmatrix}^{-1}\begin{pmatrix}
        1\\1
    \end{pmatrix}\label{Ac-4}\\
    \chi^{\mathrm{(RPA)}yy}(\mathbf{q},\omega)=\begin{pmatrix}
        -i&i
    \end{pmatrix}\chi^{(0)}(\mathbf{q},\omega)\begin{pmatrix}
        1-U\chi^{(0)}(\mathbf{q},\omega)&0\\
        0&1-U\chi^{(0)}(\mathbf{q},\omega)
    \end{pmatrix}^{-1}\begin{pmatrix}
        i\\-i
    \end{pmatrix}\\
    \chi^{\mathrm{(RPA)}zz}(\mathbf{q},\omega)=\begin{pmatrix}
        1&-1
    \end{pmatrix}\chi^{(0)}(\mathbf{q},\omega)\begin{pmatrix}
        1&U\chi^{(0)}(\mathbf{q},\omega)\\
        U\chi^{(0)}(\mathbf{q},\omega)&1
    \end{pmatrix}^{-1}\begin{pmatrix}
        1\\-1
    \end{pmatrix}
\end{align}
The charge susceptibility can also be obtained by replacing $\sigma_z$ by $\sigma_0$:
\begin{align}
    \chi^{\mathrm{(RPA)}c}(\mathbf{q},\omega)=\begin{pmatrix}
        1&1
    \end{pmatrix}\chi^{(0)}(\mathbf{q},\omega)\begin{pmatrix}
        1&U\chi^{(0)}(\mathbf{q},\omega)\\
        U\chi^{(0)}(\mathbf{q},\omega)&1
    \end{pmatrix}^{-1}\begin{pmatrix}
        1\\1
    \end{pmatrix}\label{Ac-7}
\end{align}
Simplifying Eq. \eqref{Ac-4} - \eqref{Ac-7} gives the familiar results for a normal metal
\begin{align}
    \chi^{\mathrm{(RPA)}xx}(\mathbf{q},\omega)=\chi^{\mathrm{(RPA)}yy}(\mathbf{q},\omega)&=\chi^{\mathrm{(RPA)}zz}(\mathbf{q},\omega)
    =\frac{2\chi^{(0)}(\mathbf{q},\omega)}{1-U\chi^{(0)}(\mathbf{q},\omega)}\\
    \chi^{\mathrm{(RPA)}c}(\mathbf{q},\omega)
    &=\frac{2\chi^{(0)}(\mathbf{q},\omega)}{1+U\chi^{(0)}(\mathbf{q},\omega)}
\end{align}

\section{$xx$-components of spin susceptibility}

\begin{figure}[tb]
    \centering
    \includegraphics[width=\textwidth]{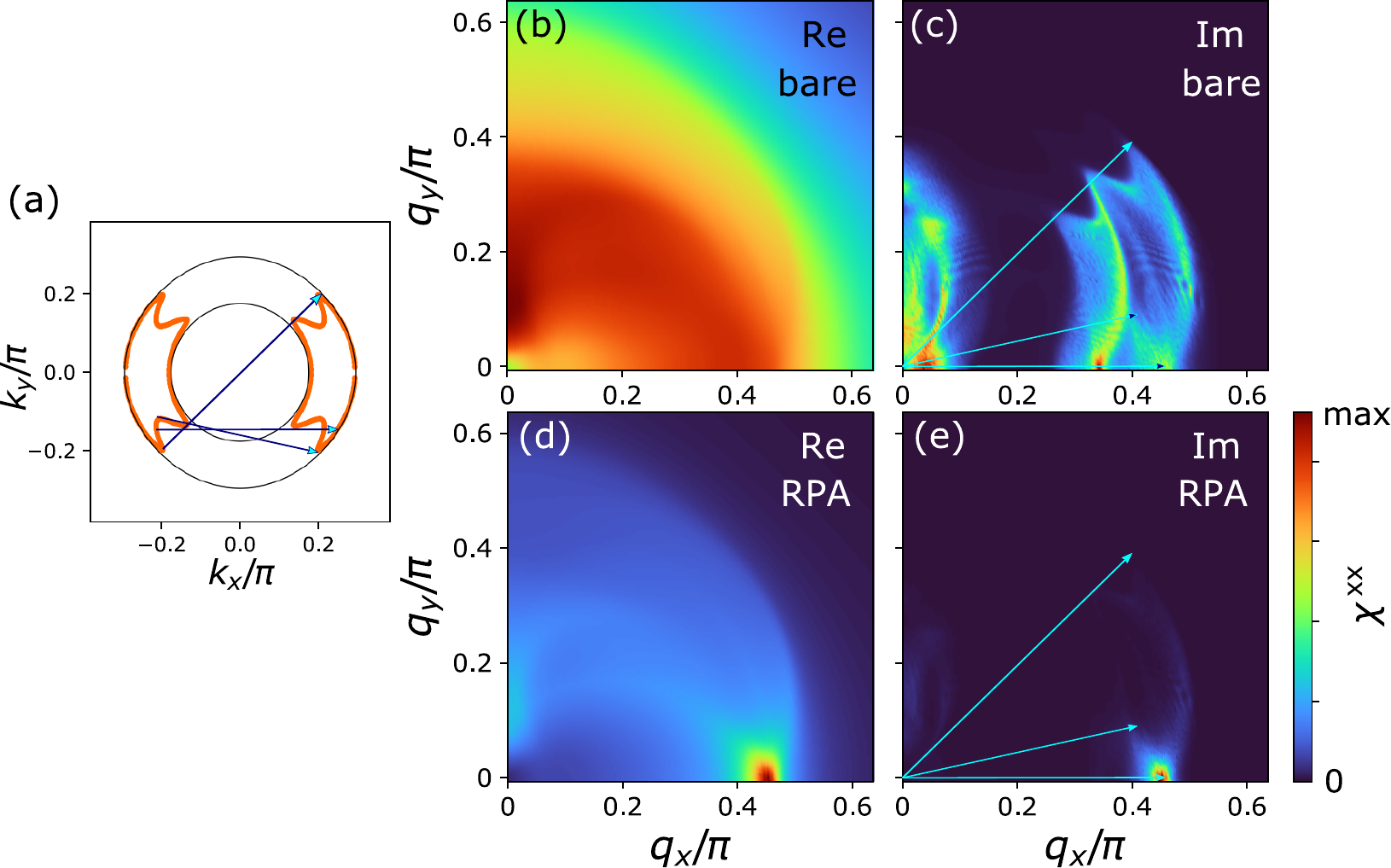}
    \caption{(a) The same Bogoliubov Fermi surface as in Fig.~2(b) of the main text. (b-e) Real and imaginary parts of $\chi^{xx}$ at $T=0$. All model parameters are the same as in Fig.~2. The arrows correspond to those in panel (a). The color bar maxima are 0.7, 0.0013, 18, 1.2 for panel (b), (c), (d), (e) respectively.}
    \label{Fig4}
\end{figure}

Here we show the same analysis to the $xx$-components of spin susceptibility as to the $zz$-components in the main text. First we check if the BFS clusters into separate groups according to the spin-flipped coherence factors $W_{l\bar l}(r\mathbf{k},s\mathbf{k'})\equiv U_{rl\mathbf{k}}U_{s\bar l\mathbf{k'}}^*$ (Eq. (9) of the main text), which is relevant to the $x$ and $y$ components of the spin susceptibility. Employing the Leiden algorithm for community detection\cite{csardi2006igraph} using the same resolution parameter, we see that the BFSs do not cluster into more than one group using linear combinations of $W_{l\bar l}$ as edge weights. Nevertheless, this only means the BFSs are more interconnected by the scattering processes relevant to the $x$ and $y$ components of the spin susceptibility, but does not necessarily mean that coherence is not important and scattering between any two points on the BFSs would equally contributes. In fact, from the $\mathbf{q}$-dependent $\chi^{xx}$ plot (Fig. \ref{Fig4} (b-d)) we see that only
the scattering between certain parts of the BFS (represented by the arrows in Fig. \ref{Fig4} (a)) contributes significantly to the $1/(T_1T)$ at large $U$ near the antiferromagnetic instability. And this is not due to a trivial enhancement from RPA since in the bare susceptibility these scattering vectors do not give particularly large contribution. These findings are all consistent with the discussion in the main text.

\begin{figure}[tb]
    \centering
    \includegraphics[width=\textwidth]{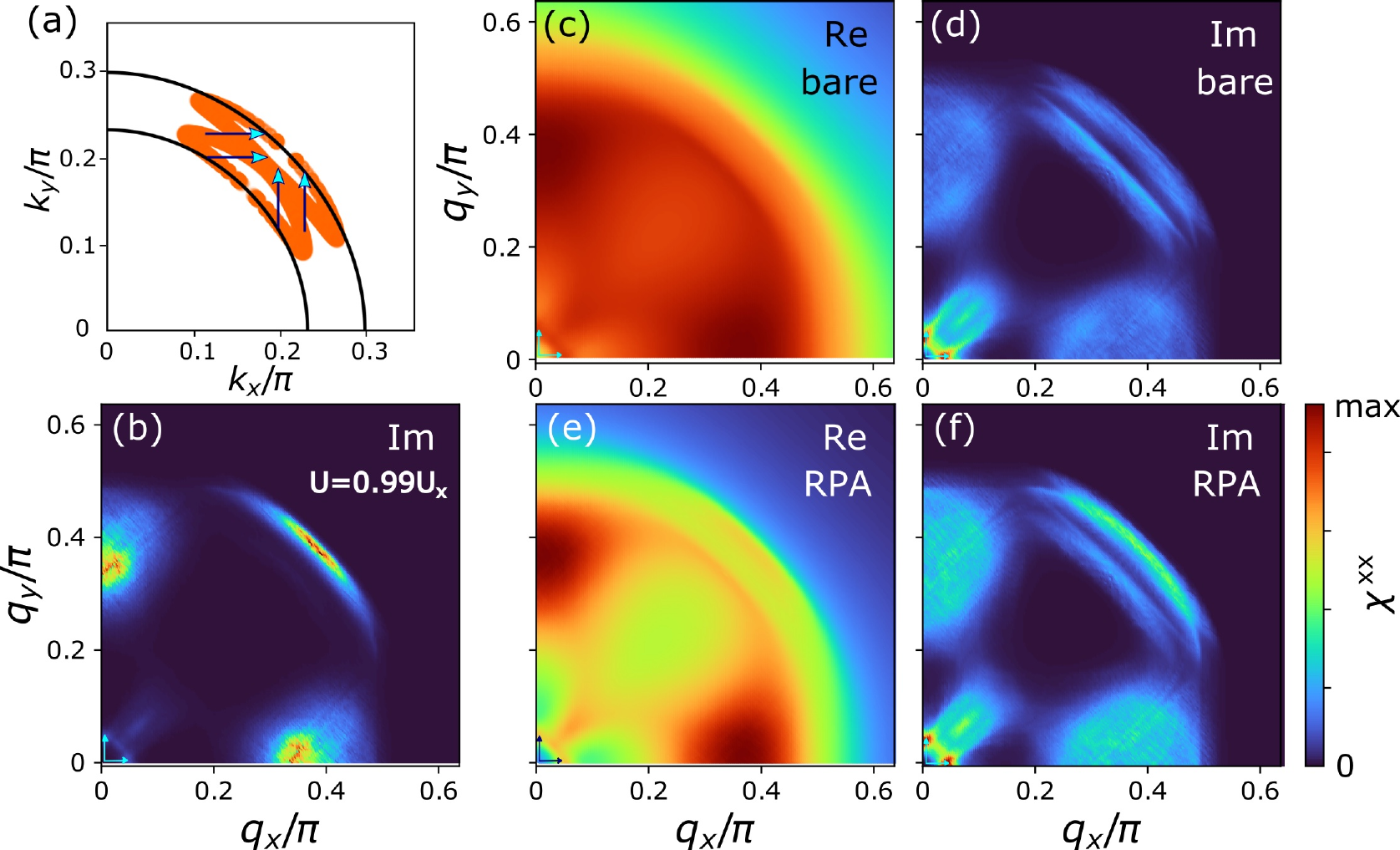}
    \caption{(a) Zoom in view of the same Bogoliubov Fermi surface as in Fig.1(b). (b) Imaginary part of $\chi^{xx}$ at $T=0$ with $U=11=0.99U_x(0)$. (c-f) Real and imaginary parts of $\chi^{xx}$ at $T=0$. All model parameters are the same as in Fig.2, including $U=10=0.99U_c(0)$. The small arrows near $\mathbf{q}=0$ correspond to those in panel (a). The color bar maxima are 6.5, 0.7, 0.006, 4.5, 0.1 for panel (b), (c), (d), (e), (f) respectively.}
    \label{Fig3}
\end{figure}

In the $C_4$ symmetric example the shift of important $\mathbf{q}$-vectors as $U$ increases still exists (compare panel (d) and (f) in Fig.~\ref{Fig3}), but is less obvious than those in Fig.~\ref{Fig4}, Figs. 1 and 2 of the main text. To better show this shifting, we plot in Fig.~\ref{Fig3} panel (b) the imaginary part of $\chi^{\text{(RPA)}xx}$ for a even larger $U$ than in panel (f). Here $U$ is larger than the critical $U$ for $\chi^{zz}$, which is also the lowest $U$ any component of the spin susceptibility could diverge, but is smaller than the critical $U$ for $\chi^{xx}$ itself. Of course Fig.~\ref{Fig3} panel (b) does not show any quantity that is physical, but it clearly shows the trend that the small $\mathbf{q}$-vectors represented by the blue arrows in panel (a), which is dominant in the bare calculation, is contributing less to the $1/(T_1T)$ as correlation gets stronger.

\bibliography{NMR.bib}